

\def\gl{\mathrel{\raise1ex\hbox{$>$\kern-.75em\lower1ex\hbox{$<$}}}}
\def\lg{\mathrel{\raise1ex\hbox{$<$\kern-.75em\lower1ex\hbox{$>$}}}}
\def\gtwid{\mathrel{\raise.3ex\hbox{$>$\kern-.75em\lower1ex\hbox{$\sim$}}}}
\def\ltwid{\mathrel{\raise.3ex\hbox{$<$\kern-.75em\lower1ex\hbox{$\sim$}}}}
\def\sqr#1#2{{\vcenter{\hrule height.#2pt
      \hbox{\vrule width.#2pt height#1pt \kern#1pt
         \vrule width.#2pt}
      \hrule height.#2pt}}}

\def\eg{\hbox{{\it e.\ g.}}}\def\ie{\hbox{{\it i.\ e.}}}\def\etc{{\it etc.}}

\def\erf{\rm erf}

\def\leaderfill{\leaders\hbox to 1em{\hss.\hss}\hfill}


\def\ref#1{${}^{#1}$}

\newcount\eqnum \eqnum=0  
\def\eqnoi{\global\advance\eqnum by 1\eqno(\the\eqnum)}
\def\eqnai{\global\advance\eqnum by 1\eqno(\the\eqnum{a})}
\def\back#1{{\advance\eqnum by-#1 Eq.~(\the\eqnum)}}
\def\last{Eq.~(\the\eqnum)}                   

\newcount\refnum\refnum=0  
\def\refi{\smallskip\global\advance\refnum by 1\item{\the\refnum.}}

\newcount\rfignum\rfignum=0  
\def\rfigi{\medskip\global\advance\rfignum by 1\item{Figure \the\rfignum.}}

\newcount\fignum\fignum=0  
\def\figi{\global\advance\fignum by 1 Fig.~\the\fignum}

\newcount\rtabnum\rtabnum=0  
\def\rtabi{\medskip\global\advance\rtabnum by 1\item{Table \the\rtabnum.}}

\newcount\tabnum\tabnum=0  
\def\tabi{\global\advance\tabnum by 1 Table~\the\tabnum}

\newcount\secnum\secnum=0 
\def\chap#1{\global\advance\secnum by 1
\bigskip\centerline{\bf{\the\secnum}. #1}\smallskip\noindent}
\def\subsec#1{\smallskip{\bf\centerline{#1}\smallskip}}

\def\pd#1#2{{\partial #1\over\partial #2}}      
\def\p2d#1#2{{\partial^2 #1\over\partial #2^2}} 
\def\t2d#1#2{{d^2 #1\over d #2^2}} 
\def\av#1{\langle #1\rangle}                    

\def\2kth{{$2k^{\rm th}$}}

\def\n-th{{$(n-1)^{\rm th}$}}

\def\N-th{{$(N-1)^{\rm th}$}}

\def\0th{$0^{\rm th}$}
\def\1st{$1^{\rm st}$}
\def\2nd{$2^{\rm nd}$}
\def\3rd{$3^{\rm rd}$}
\def\4th{$4^{\rm th}$}
\def\5th{$5^{\rm th}$}
\def\5th{$6^{\rm th}$}
\def\6th{$7^{\rm th}$}
\def\7th{$7^{\rm th}$}
\def\8th{$8^{\rm th}$}
\def\9th{$9^{\rm th}$}

\def\aap #1 #2 #3 {{\sl Adv.\ Appl.\ Prob.} {\bf #1}, #2 (#3)}
\def\ac #1 #2 #3 {{\sl Adv.\ Catal.} {\bf #1}, #2 (#3)}
\def\aces #1 #2 #3 {{\sl Adv.\ Chem.\ Eng.\ Sci.} {\bf #1}, #2 (#3)}
\def\acp #1 #2 #3 {{\sl Adv.\ Chem.\ Phys.} {\bf #1}, #2 (#3)}
\def\ae #1 #2 #3 {{\sl Ann.\ Eugenics} {\bf #1}, #2 (#3)}
\def\ah #1 #2 #3 {{\sl Adv.\ Hydrosci.} {\bf #1}, #2 (#3)}
\def\aj #1 #2 #3 {{\sl Astrophys.\ J.} {\bf #1}, #2 (#3)}
\def\ajp #1 #2 #3 {{\sl Am.\ J.\ Phys.} {\bf #1}, #2 (#3)}
\def\ams #1 #2 #3 {{\sl Ann.\ Math.\ Stat.} {\bf #1}, #2 (#3)}
\def\ap #1 #2 #3 {{\sl Adv.\ Phys.} {\bf #1}, #2 (#3)}
\def\apa #1 #2 #3 {{\sl Appl.\ Phys.\ A} {\bf #1}, #2 (#3)}
\def\apc #1 #2 #3 {{\sl Appl.\ Catal.} {\bf #1}, #2 (#3)}
\def\annp #1 #2 #3 {{\sl Ann.\ Phys.\ (N.Y.)} {\bf #1}, #2 (#3)}
\def\annpr #1 #2 #3 {{\sl Ann.\ Probab.} {\bf #1}, #2 (#3)}
\def\arfm #1 #2 #3 {{\sl Ann.\ Rev.\ Fluid Mech.} {\bf #1}, #2 (#3)}
\def\arpc #1 #2 #3 {{\sl Ann.\ Rev.\ Phys.\ Chem.} {\bf #1}, #2 (#3)}
\def\ast #1 #2 #3 {{\sl Aerosol Sci.\ Tech.} {\bf #1}, #2 (#3)}
\def\bcdg #1 #2 #3 {{\sl Ber.\ Chem.\ Dtsch.\ Ges.} {\bf #1}, #2 (#3)}
\def\bio #1 #2 #3 {{\sl Biometrika} {\bf #1}, #2 (#3)}
\def\bj #1 #2 #3 {{\sl Biophys.\ J.} {\bf #1}, #2 (#3)}
\def\ces #1 #2 #3 {{\sl Chem.\ Engr.\ Sci.} {\bf #1}, #2 (#3)}
\def\cf #1 #2 #3 {{\sl Combust.\ and Flame} {\bf #1}, #2 (#3)}
\def\cp #1 #2 #3 {{\sl Chem.\ Phys.} {\bf #1}, #2 (#3)}
\def\cpam #1 #2 #3 {{\sl Commun.\ Pure Appl.\ Math.} {\bf #1}, #2 (#3)}
\def\crev #1 #2 #3 {{\sl Catal.\ Rev.} {\bf #1}, #2 (#3)}
\def\eul #1 #2 #3 {{\sl Europhys.\ Lett.} {\bf #1}, #2 (#3)}
\def\ic #1 #2 #3 {{\sl Icarus} {\bf #1}, #2 (#3)}
\def\iec #1 #2 #3 {{\sl Ind.\ Eng.\ Chem.} {\bf #1}, #2 (#3)}
\def\ijf #1 #2 #3 {{\sl Int.\ J.\ of Fracture} {\bf #1}, #2 (#3)}
\def\ijmp #1 #2 #3 {{\sl Int.\ J.\ Modern Phys.} {\bf #1}, #2 (#3)}
\def\ijss #1 #2 #3 {{\sl Int.\ J.\ Solids Structures} {\bf #1}, #2 (#3)}
\def\jam #1 #2 #3 {{\sl J.\ Appl.\ Mech.} {\bf #1}, #2 (#3)}
\def\jap #1 #2 #3 {{\sl J.\ Appl.\ Phys.} {\bf #1}, #2 (#3)}
\def\japr #1 #2 #3 {{\sl J.\ Appl.\ Prob.} {\bf #1}, #2 (#3)}
\def\jaes #1 #2 #3 {{\sl J.\ Aerosol.\ Sci.} {\bf #1}, #2 (#3)}
\def\jas #1 #2 #3 {{\sl J.\ Atmos.\ Sci.} {\bf #1}, #2 (#3)}
\def\jc #1 #2 #3 {{\sl J.\ Catal.} {\bf #1}, #2 (#3)}
\def\jcis #1 #2 #3 {{\sl J.\ Colloid Interface Sci.} {\bf #1}, #2 (#3)}
\def\jcp #1 #2 #3 {{\sl J.\ Chem.\ Phys.} {\bf #1}, #2 (#3)}
\def\jdep #1 #2 #3 {{\sl J.\ de Physique I} {\bf #1}, #2 (#3)}
\def\jdepl #1 #2 #3 {{\sl J. de Physique Lett.} {\bf #1}, #2 (#3)}
\def\jetp #1 #2 #3 {{\sl Sov.\ Phys.\ JETP} {\bf #1}, #2 (#3)}
\def\jetpl #1 #2 #3 {{\sl Sov. Phys.\ JETP Letters} {\bf #1}, #2 (#3)}
\def\jfi #1 #2 #3 {{\sl J. Franklin Inst.} {\bf #1}, #2 (#3)}
\def\jfm #1 #2 #3 {{\sl J. Fluid Mech.} {\bf #1}, #2 (#3)}
\def\jgr #1 #2 #3 {{\sl J.\ Geophys.\ Res.} {\bf #1}, #2 (#3)}
\def\jif #1 #2 #3 {{\sl J. Inst.\ Fuel} {\bf #1}, #2 (#3)}
\def\jmo #1 #2 #3 {{\sl J. Mod. Opt.} {\bf #1}, #2 (#3)}
\def\jmp #1 #2 #3 {{\sl J. Math. Phys.} {\bf #1}, #2 (#3)}
\def\jms #1 #2 #3 {{\sl J. Memb.\ Sci.} {\bf #1}, #2 (#3)}
\def\josaA #1 #2 #3 {{\sl J. Opt.\ Soc.\ Am.\ A} {\bf #1}, #2 (#3)}
\def\josaB #1 #2 #3 {{\sl J. Opt.\ Soc.\ Am.\ B } {\bf #1}, #2 (#3)}
\def\jpa #1 #2 #3 {{\sl J. Phys.\ A} {\bf #1}, #2 (#3)}
\def\jpc #1 #2 #3 {{\sl J. Phys.\ C} {\bf #1}, #2 (#3)}
\def\jpchem #1 #2 #3 {{\sl J. Phys.\ Chem.} {\bf #1}, #2 (#3)}
\def\jps #1 #2 #3 {{\sl J. Polymer.\ Sci.} {\bf #1}, #2 (#3)}
\def\jpsj #1 #2 #3 {{\sl J. Phys.\ Soc. Jpn.} {\bf #1}, #2 (#3)}
\def\jsp #1 #2 #3 {{\sl J. Stat.\ Phys.} {\bf #1}, #2 (#3)}
\def\jtb #1 #2 #3 {{\sl J. Theor.\ Biol.} {\bf #1}, #2 (#3)}
\def\kc #1 #2 #3 {{\sl Kinet.\ Catal.\ (USSR)} {\bf #1}, #2 (#3)}
\def\macro #1 #2 #3 {{\sl Macromolecules} {\bf #1}, #2 (#3)}
\def\mclc #1 #2 #3 {{\sl Mol.\ Cryst.\ Liq.\ Cryst.} {\bf #1}, #2 (#3)}
\def\mubm #1 #2 #3 {{\sl Moscow Univ.\ Bull.\ Math.} {\bf #1}, #2 (#3)}
\def\nat #1 #2 #3 {{\sl Nature} {\bf #1}, #2 (#3)}
\def\npa #1 #2 #3 {{\sl Nucl.\ Phys.\ A} {\bf #1}, #2 (#3)}
\def\npb #1 #2 #3 {{\sl Nucl.\ Phys. B} {\bf #1}, #2 (#3)}
\def\PA #1 #2 #3 {{\sl PAGEOPH} {\bf #1}, #2 (#3)}
\def\pA #1 #2 #3 {{\sl Physica A} {\bf #1}, #2 (#3)}
\def\pBC #1 #2 #3 {{\sl Physica B \& C} {\bf #1}, #2 (#3)}
\def\pD #1 #2 #3 {{\sl Physica D} {\bf #1}, #2 (#3)}
\def\pams #1 #2 #3 {{\sl Proc.\ Am.\ Math.\ Soc.} {\bf #1}, #2 (#3)}
\def\pcps #1 #2 #3 {{\sl Proc.\ Camb.\ Philos.\ Soc.} {\bf #1}, #2 (#3)}
\def\pf #1 #2 #3 {{\sl Phys.\ Fluids} {\bf #1}, #2 (#3)}
\def\pla #1 #2 #3 {{\sl Phys.\ Lett. A} {\bf #1}, #2 (#3)}
\def\plb #1 #2 #3 {{\sl Phys.\ Lett. B} {\bf #1}, #2 (#3)}
\def\pmB #1 #2 #3 {{\sl Philos.\ Mag. B} {\bf #1}, #2 (#3)}
\def\pnas #1 #2 #3 {{\sl Proc.\ Natl.\ Acad.\ Sci.} {\bf #1}, #2 (#3)}
\def\pr #1 #2 #3 {{\sl Phys.\ Rev.} {\bf #1}, #2 (#3)}
\def\pra #1 #2 #3 {{\sl Phys.\ Rev.\ A} {\bf #1}, #2 (#3)}
\def\prb #1 #2 #3 {{\sl Phys.\ Rev.\ B} {\bf #1}, #2 (#3)}
\def\prc #1 #2 #3 {{\sl Phys.\ Rev.\C} {\bf #1}, #2 (#3)}
\def\prd #1 #2 #3 {{\sl Phys.\ Rev.\ D} {\bf #1}, #2 (#3)}
\def\pre #1 #2 #3 {{\sl Phys.\ Rev.\ E} {\bf #1}, #2 (#3)}
\def\prept #1 #2 #3 {{\sl Phys.\ Repts.} {\bf #1}, #2 (#3)}
\def\prk #1 #2 #3 {{\sl Prog.\ Reac.\ Kinetics} {\bf #1}, #2 (#3)}
\def\prl #1 #2 #3 {{\sl Phys.\ Rev.\ Lett.} {\bf #1}, #2 (#3)}
\def\prsl #1 #2 #3 {{\sl Proc.\ Roy.\ Soc.\ London Ser. A} {\bf #1}, #2 (#3)}
\def\pss #1 #2 #3 {{\sl Prog.\ Surf.\ Sci.} {\bf #1}, #2 (#3)}
\def\pt #1 #2 #3 {{\sl Powder Tech.} {\bf #1}, #2 (#3)}
\def\ptrs #1 #2 #3 {{\sl Phil.\ Trans.\ Roy.\ Soc., Ser. A} {\bf #1}, #2 (#3)}
\def\rjpc #1 #2 #3 {{\sl Russ.\ J. Phys.\ Chem.} {\bf #1}, #2 (#3)}
\def\rmp #1 #2 #3 {{\sl Rev.\ Mod.\ Phys.} {\bf #1}, #2 (#3)}
\def\rpp #1 #2 #3 {{\sl Rep.\ Prog.\ Phys.} {\bf #1}, #2 (#3)}
\def\sci #1 #2 #3 {{\sl Science} {\bf #1}, #2 (#3)}
\def\sciam #1 #2 #3 {{\sl Scientific American} {\bf #1}, #2 (#3)}
\def\SIAMjam #1 #2 #3 {{\sl SIAM J.\ Appl.\ Math.} {\bf #1}, #2 (#3)}
\def\spu #1 #2 #3 {{\sl Sov.\ Phys.\ Usp.} {\bf #1}, #2 (#3)}
\def\SPEre #1 #2 #3 {{\sl SPE Reservoir Eng.} {\bf #1}, #2 (#3)}
\def\ssc #1 #2 #3 {{\sl Sol.\ State.\ Commun.} {\bf #1}, #2 (#3)}
\def\ss #1 #2 #3 {{\sl Surf.\ Sci.} {\bf #1}, #2 (#3)}
\def\stec #1 #2 #3 {{\sl Sov.\ Theor.\ Exp.\ Chem.} {\bf #1}, #2 (#3)}
\def\tAIME #1 #2 #3 {{\sl Trans.\ AIME} {\bf #1}, #2 (#3)}
\def\tfs #1 #2 #3 {{\sl Trans.\ Faraday Soc.} {\bf #1}, #2 (#3)}
\def\tpa #1 #2 #3 {{\sl Theor.\ Prob.\ Appl.} {\bf #1}, #2 (#3)}
\def\tpm #1 #2 #3 {{\sl Transport in Porous Media} {\bf #1}, #2 (#3)}
\def\usp #1 #2 #3 {{\sl Sov.\ Phys. -- Usp.} {\bf #1}, #2 (#3)}
\def\wrr #1 #2 #3 {{\sl Water Resources Res.} {\bf #1}, #2 (#3)}
\def\zpb #1 #2 #3 {{\sl Z. Phys.\ B} {\bf #1}, #2 (#3)}
\def\zpc #1 #2 #3 {{\sl Z. Phys.\ Chem.} {\bf #1}, #2 (#3)}
\def\ztf #1 #2 #3 {{\sl Zh. Tekh.\ Fiz.} {\bf #1}, #2 (#3)}
\def\zw #1 #2 #3 {{\sl Z. Wahrsch.\ verw.\ Gebiete} {\bf #1}, #2 (#3)}

\magnification=1200 \baselineskip=14truebp
\newcount\secnum\newcount\eqnum\newcount\subnum\eqnum=0\secnum=0\subnum=0
\def\eqnoi{\global\advance\eqnum by 1\eqno(\the\secnum.\the\eqnum)}
\def\chapter#1{\global\advance\secnum by
1\eqnum=0\subnum=0\bigskip\bigskip\centerline{\bf{\the\secnum}:
#1}\bigskip\medskip\noindent}
\def\subsec#1{\global\advance\subnum by 1\medskip
{\bf\centerline{\the\secnum.\the\subnum: #1}\medskip\noindent}}
\def\back#1{{\advance\eqnum by-#1  Eq.~(\the\secnum.\the\eqnum)}}
\def\last{Eq.~(\the\secnum.\the\eqnum)}
\def\eqnai{\global\advance\eqnum by 1\eqno(\the\secnum.\the\eqnum{a})}
\def\part#1#2{{\partial#1\over\partial#2}}

\def\corrone#1{\rho_#1}\def\corrtwo#1#2{\rho_{#1,#2}}
\def\corrthree#1#2#3{\rho_{#1,#2,#3}}

\centerline{\bf Propagation and Extinction in Branching Annihilating Random
Walks}
\bigskip
\centerline{D. ben-Avraham$\dag$, F. Leyvraz$\ddag$, and S. Redner$\star$}
\bigskip
\centerline{$\dag$Clarkson Institute for Statistical Physics (CISP) and
Department of Physics,}
\centerline{Clarkson University, Potsdam, NY~13699-5820}
\medskip
\centerline{$\ddag$Instituto de Fisica, Laboratorio de Cuernavaca, UNAM}
\centerline{Apdo Postal 20-364, 01000 Mexico D. F., MEXICO}
\medskip
\centerline{$\star$Center for Polymer Studies and Department of Physics}
\centerline{Boston University, Boston, MA~ 02215}
\bigskip
\noindent
We investigate the temporal evolution and spatial propagation of
branching annihilating random walks in one dimension.  Depending on the
branching and annihilation rates, a few-particle initial state can
evolve to a propagating finite density wave, or extinction may occur, in
which the number of particles vanishes in the long-time limit.  The
number parity conserving case where 2-offspring are produced in each
branching event can be solved exactly for unit reaction probability,
from which qualitative features of the transition between propagation
and extinction, as well as intriguing parity-specific effects are
elucidated.  An approximate analysis is developed to treat this
transition for general BAW processes.  A scaling description suggests
that the critical exponents which describe the vanishing of the particle
density at the transition are unrelated to those of conventional models,
such as Reggeon Field Theory.
\bigskip
\noindent P. A. C. S. Numbers: 02.50.+s, 05.40.+j, 82.20.-w
\bigskip

\chapter{Introduction}%
In the branching annihilating random walk (BAW), a single
random walk branches at some specified rate and two random walkers
annihilate at another rate when they meet.\ref{1-4}~ As a function of
these rates, the number of random walkers may grow without bound,
reach a finite limiting number, or vanish asymptotically.  Our goal,
in this paper, is to determine some of the long-time properties of
this BAW process.  We are particularly interested in understanding the
kinetics and density distribution when the initial state consists of a
small number of localized particles.

Interest in this process has several motivations.  First, considerable
theoretical effort has been devoted to establishing the existence of and
quantifying the non-equilibrium phase transition between ``propagation''
and ``extinction'' for a variety of interacting particle systems\ref{5}
which are closely related to BAWs.  Here the term extinction refers to
the situation where annihilation dominates over branching and an
initially localized population of particles ultimately disappears.  In
the complementary case, branching dominates over annihilation and an
initially localized population evolves into a propagating wavefront
which advances into the otherwise empty system.  Typical examples of
these phenomena include the contact process,\ref{6} Schl\"ogl
models,\ref{7} as well as directed percolation and Reggeon field
theory.\ref{8}~ The propagation phenomenon is also a discrete
realization of the ``Fisher wave'',\ref{9,10} which describes the
continuum dynamics of an initially localized single-species population
whose evolution is influenced by diffusion, as well as by both (linear)
birth and (quadratic) death mechanisms.  The relation between the
continuum description of the Fisher wave and the corresponding BAW is
tenuous and comprehensive investigations of discrete BAW models would be
helpful to understand better the relation with the continuum
counterpart.  Second, there is a direct correspondence between the
2-offspring BAW and the interface dynamics in the reaction-limited
monomer-monomer surface reaction model in one dimension.\ref{11,12}~ For
the surface reaction, a lattice is filled with $A$ and $B$ particles and
the ensuing dynamics is defined by randomly and repeatedly selecting a
nearest-neighbor $AB$ pair and changing it to either $AA$, $BB$, $AB$,
or $BA$ at specified rates.  The dynamics of $AB$ interfaces is
identical to that of the individual particles in the 2-offspring BAW.

In addition to connections with various non-equilibrium systems, the BAW
model is amenable to theoretical analysis.  Somewhat surprisingly, we
find that the transition between propagation and extinction is
controlled by detailed features of the underlying discrete process.  In
particular, the exact solution of the 2-offspring BAW model in one
dimension reveals that propagation occurs only for infinite branching
rate and the parity of the initial number of particles essentially
influences the long-time kinetics.

In the next section, we outline several basic facts about the BAW
process.  The general conditions which lead to propagation or
extinction are discussed. In Section 3, we present an exact solution
for the evolution of the 2-offspring BAW process $A\to3A$ and
$2A\to0$, in the case where the probability of reaction when two
particles meet, $k$, is unity.  For this case, extinction is
paradoxically found to occur for all non-zero values of the diffusion
rate.  When there is only branching, a behavior intermediate to
propagation and extinction occurs.  In Section 4, we present an
approximate description for the transition between propagation and
extinction by solving a truncated hierarchy of rate equations for
multi-particle correlation functions.  The primary result of this
treatment is that the transition emerges naturally at the next level
of approximation beyond mean-field theory.  In Section 5, we present
numerical simulation results to support our various theoretical
predictions, and then conclude in Section 6.

\chapter{Models and Phenomenology}%
We define the branching annihilating random walk (BAW) on a lattice as
follows.  A particle is picked at random.  It can undergo either
nearest-neighbor diffusion or it may branch with respective rates $D$
and $r$.  In a diffusion attempt, a random direction is picked and the
particle moves to the target site if it is unoccupied.  If the target
site is occupied, then annihilation of the incident and target particles
occurs with probability $k$.  Otherwise, the incident particle remains
at its original position.  The details of the branching step depend on
the number of offspring produced.  In the 2-offspring BAW, if a
branching attempt is selected (at rate $r$), then branching to the two
nearest-neighbor sites occurs with probability $1$ if both neighbors are
empty.  If one or both neighbors are occupied, then branching to both
neighbors occurs with probability $k$.  In this branching, if a newly
created particle is placed on a previously occupied site, then both
particles are removed.  An analogous procedure is employed for the
1-offspring process.  Namely, in a branching event, if the target is
occupied, then branching occurs with probability $k$.  This branching
entails immediate annihilation of the newly-created and target
particles.  While these microscopic rules are somewhat involved when
both branching and a finite reaction rate are operative, any reasonable
discrete realization of the continuum process is anticipated to give
qualitatively similar results.  (An exception is the case of parallel
dynamics, as opposed to the serial updating assumed here.)~ Since $D$
and $r$ have dimensions of inverse time and $k$ is dimensionless, two
dimensionless parameters which characterize the system are $D/r$ and
$k$.

As mentioned above, the BAW process with an initially localized
particle population generically exhibits a transition from extinction
(which is of one of two types; see below) to propagation.  However,
this transition appears to be at odds with a standard mean-field
treatment of the model.  Indeed, such a description of the BAW
leads to the reaction-diffusion equation
$$
\pd{\rho(\vec r,t)}{t}=a\rho(\vec r,t)-b\rho(\vec r,t)^2+D\Delta
\rho(\vec r,t),\eqnoi
$$
where $\rho(\vec r,t)$ is the particle density at position $\vec r$ at
time $t$.  Here $a$, $b$, and $D$ correspond roughly to the branching
rate, the reaction rate, and the diffusion rate, respectively.  This
equation admits a propagating Fisher wave solution of asymptotic
density $\rho_\infty=a/b$ which advances into the low-density region
with a velocity proportional to $\sqrt{aD}$, whenever $a>0$.  Since
$a$ is implicitly positive in this description, there appears to be no
mechanism for extinction.

To generate a transition to extinction, a mechanism which changes the
sign of $a$ is needed.  Such a mechanism, however, is easily realizable.
At low densities, mean-field theory erroneously postulates that the
encounter probability varies as the square of the particle density,
since all spatial correlations are neglected.  However, a more complete
treatment of correlations leads to an encounter probability which is
{\it linear} in the particle density in the low density limit.  This
follow from the exact reaction-diffusion equation
$$
\pd{\rho(\vec r,t)}{t}=a_{\rm exact}\rho(\vec r,t)-b_{\rm exact}\rho_2(\vec
0,t)+D\Delta
\rho(\vec r,t),\eqnoi
$$
where $\rho_2(0,t)$ denotes the two-point correlation function at zero
separation, and branching is accounted for by $a_{\rm exact}>0$.
However, for $\rho\ll\sqrt{a_{\rm exact}/D}$, any reasonable theory
must give $\rho_2(\vec 0,t)\simeq\av{\rho} +\av{\rho}^2$ because the
birth mechanism produces nearby {\it pairs} of particles.  Thus,
extinction may arise in low spatial dimension because
branching and annihilation terms are of comparable magnitude in
\back1; in particular, the sign of the coarse-grained branching rate
$a\equiv a_{\rm exact}-b_{\rm exact}$ may change as external
parameters are varied.  In the following, we shall use $r$ for the
branching rate and $k$ for the annihilation probability.

In numerical simulations, a transition from propagation to extinction
generically occurs as the ratio of diffusion to branching increases
(Fig.~1).  As the annihilation probability (equivalent to the reaction
rate) increases, the extinction transition is shifted to lower values of
$D/r$.  A crucial determinant for extinction is the ``parity'' of the
branching event.  For example, an even-offspring BAW process which
starts with an odd number of particles cannot exhibit true extinction.
Rather, the number of particles remains bounded when $D/r$ is
large.  Thus extinction comes in two versions: In non-parity preserving
models (odd-offspring BAW), or in parity preserving models with an even
number of particles initially, there is true extinction, whereas in the
complementary cases, the particle number remains bounded.

The special case of the 2-offspring BAW in the limits of pure branching,
$D/r=0$, and also complete reaction, $k=1$, deserves emphasis because of
its peculiar and easily visualized features (Fig.~2).  The crucial
aspect of this situation is that a pair of nearest-neighbor particles
diffuses rigidly under the action of the branching process.  Through the
branching process, a single initial particle spawns rigid pairs at a
finite rate which then diffuse freely.  In this situation, the number of
particles grows as $\sqrt{t}$.  It is only when there is non-zero
diffusion that extinction can occur.

\chapter{Exact Solution for the 2-Offspring BAW in One Dimension}%

In this section, we derive the exact solution for the one-dimensional
2-offspring BAW in the case of unit reaction probability.  This
provides a complete description of the transition between extinction
and propagation.  Our approach is based on solving for the time
evolution of $P_{j,k}(t)$, the probability that there is an odd number
of particles in the interval $[j,k]$.  (Notice that the particle
density at site $j$ is simply $P_{j,j}(t)$.)~ In a spatially
homogeneous system, $P_{j,k}(t)$ is a function of $j-k$ only and
satisfies a soluble single-variable diffusion equation\ref{12} while
for the heterogeneous system, $P_{j,k}(t)$ is a function of both $j$
and $k$.  In the continuum limit, the master equation for $P_{j,k}(t)$
can be written as a 2-dimensional diffusion equation with a radiation
boundary condition.

To write the master equation for $P_{j,k}(t)$, define $D_0$ and $r_0$ as
the hopping and branching rates, respectively, in a elemental event.  In
a single time step, the parity of the number of particles within $[j,k]$
can change only by hopping or branching events at $j-1$, $j$, $k$, or
$k+1$.  From straightforward but somewhat tedious bookkeeping of all
these microscopic processes (Fig.~3), we obtain the following equations
for $P_{j,k}$ for $j\geq k+1$,
$$
P_{j,k}(t+1/N)-P_{j,k}(t)={D_0+r_0\over
N}(P_{j+1,k}(t)+P_{j,k+1}(t) +P_{j,k-1}(t)+P_{j-1,k}(t)-4P_{j,k}(t)),\eqnoi
$$
while for $j=k$,
$$ P_{j,j}(t+1/N)-P_{j,j}(t)={D_0+r_0\over N}
(P_{j+1,j}(t)+P_{j,j+1}(t) -2P_{j,j}(t))-{2D_0\over N}P_{j,j}(t).\eqnoi
$$
The first equation can be extended to include $j=k$ by
introducing the boundary condition,
$$
(D_0+r_0)(P_{j,j-1}(t)+P_{j+1,j}(t))=2r_0P_{j,j}(t).\eqnoi
$$
Here we have taken the time increment for an individual birth or
diffusion event to be of the order of the inverse system size so
that each particle will typically be updated once in a unit
time step.

These equations can be simplified by transforming to the coordinates
$x=j+k$ and $y=j-k$.  After taking continuous
time derivatives, one finds, for $y\geq 1$,
$$
\dot P_{x,y}(t)=(D_0+r_0)(P_{x+1,y+1}(t)+P_{x+1,y-1}(t)+P_{x-1,y+1}(t)+
P_{x-1,y-1}(t)-4P_{x,y}(t)),\eqnoi
$$
while for $y=0$ one has,
$$
\dot
P_{x,0}(t)=(D_0+r_0)(P_{x+1,1}(t)+P_{x-1,1}(t)-2P_{x,0}(t))-2D_0P_{x,0}(t).
\eqnoi
$$
These two equations can again be unified by imposing the boundary condition,
$$
(D_0+r_0)(P_{x+1,-1}(t)+P_{x-1,-1}(t)-2P_{x,0}(t))=-2D_0P_{x,0}(t).\eqnoi
$$

To obtain a non-singular continuum limit for these master equations,
$(D_0+r_0)$ must be of order $(\Delta x)^{-2}$, whereas from the
boundary condition, $(D_0+r_0)/D_0$ should be of order $(\Delta
x)^{-1}$.  These two restrictions, in turn, imply that $r_0\sim(\Delta
x)^{-2}$ and $D_0\sim(\Delta x)^{-1}$.  That is, an infinitely large
microscopic birth rate is needed to counteract the instant
recombination of newly-formed pairs in the continuum limit, thus
ensuring a finite continuum birth rate.  In the following, we simply
replace the microscopic combination $D_0+r_0$ by $r_0$---since $r_0$
is infinitely larger than $D_0$---and then take the continuum limit.
Denoting the continuum limits of $(\Delta x)D_0$ and $(\Delta x)^2r_0$
as $D$ and $r$, respectively, the equation of motion for the interval
occupation probability $P(x,y)$ becomes
$$
\part{P(x,y)}{t}=r\nabla^2 P(x,y),\eqnoi
$$
with the corresponding boundary condition
$$
\part{P(x,y)}{y}\vert_{y=0}={D\over r}P(x,y)\vert_{y=0}.\eqnoi
$$
When $D/r=0$, a slightly different line of reasoning must be used to
determine the equation of motion, but the diffusion equation given
above remains valid.

This equation can be solved by writing the solution as a
superposition of incoming and outgoing modes in $y$ in Fourier space,
and imposing the boundary condition to relate the amplitudes of
the incoming and outgoing modes.  Upon inverting the Fourier transform,
one finds
$$
P(x,y;x_0,y_0;t)=\int{dk_1\,dk_2\over(2\pi)^2}
e^{ik_1(x-x_0)-r(k_1^2+k_2^2)t}\left(e^{ik_2(y-y_0)}
+{k_2+iD/r\over k_2-iD/r}e^{-ik_2(y+y_0)}\right),\eqnoi
$$
where $P(x,y;x_0,y_0;t)$ denotes the propagator from $(x_0, y_0)$ to
$(x,y)$ in time $t$.  For computing the total number of particles, we
only need the propagator with final spatial co-ordinate
$y=0$.  In the long time limit, this restricted propagator has the form
$$
P(x,0;x_0,y_0;t)\approx{y_0\over4\pi Drt^{3/2}}\exp\{
-{(x-x_0)^2+y_0^2 \over 4rt}\},\eqnoi
$$
which arises by neglecting the dimensionless ``radiation'' length $r/D$
compared to the diffusion length $\sqrt{Dt}$.

For a single particle present initially at the origin, $P(x_0,y_0;t=0)$
is zero for $y_0<|x_0|$ and is equal to one for $y_0\geq|x_0|$.  This
corresponds to $P(x_0,y_0;t=0)=\Theta(y_0-|x_0|)$, where $\Theta(x)$ is the
Heaviside step function.  Thus for the density distribution we find
$$
\eqalign{
c(x,t)&=\int_0^\infty dy_0\int_{-y_0}^{y_0}dx_0\,P(x,0;x_0,y_0t),\cr
&={e^{-x^2/(8rt)}\over (D/r)\sqrt{2\pi r t}}.}\eqnoi
$$
For the case of two particles which are originally separated by a
distance $d$, $P(x_0,y_0;t=0)=1$ when $|\,|x_0|-y_0|<d$, and $P(x_0,y_0;t=0)=0$
otherwise.  This initial condition leads to an expression of a similar
form to \last, except that there is an additional multiplicative factor
of $d/\sqrt{rt}$.  Thus the total number of particles,
$N(t)\equiv\int{c(x,t)\,dx}\to 2r/D$ as $t\to\infty$ when one particle is
initially present, and $N(t)\sim (rt)^{-1/2}$ when two particles are
initially present.  Generally, an initially localized group with an odd number
of particles leads to a finite number of particles as $t\to\infty$,
while a localized initial population with an even number of particles
leads to $N(t)\propto t^{-1/2}$.

If there is no diffusion, then \back3\ reduces to a Neumann boundary
condition.  With this simplification, one finds the following density
distribution for a single particle initial condition, after several
straightforward steps and without any approximations other than those
involved in the continuum limit,
$$
\eqalign{
c(x,t)&=\int_0^\infty dy_0\int_{-y_0}^{y_0}dx_0P(x,0;x_0,y_0;t),\cr
&={1\over4}\left(1-\erf^2\left({x\over2\sqrt{rt}}\right)\right).\cr }\eqnoi
$$
Thus for a single particle initial condition, $N(t)\propto\sqrt t$,
intermediate to the limiting cases of Fisher wave propagation, where the
particle number grows linearly in $t$, and extinction, where the
particle number either remains constant (through parity effects) or else
decays as $t^{-1/2}$.  Another interesting feature of the distribution
is that it qualitatively resembles a Gaussian.  This can be made
plausible by consideration of the underlying discrete process.  When
there is only branching, it is easy to verify that a nearest-neighbor
particle pair propagates diffusively as a rigid unit (Fig.~2).  When two
such soliton-like excitations meet, they merely ``scatter'' without any
change in their form.  As a single initial particle evolves, there is
production of pairs at a finite rate which then diffuse freely within
a region of length $\sqrt{rt}$.  Conversely for an even number of
particles initially, ``interference'' between the offspring of the
initial seed particles leads to the density vanishing as $t^{-1/2}$ in
the long-time limit.

When the reaction rate $k<1$, then the transition between extinction
and Fisher wave propagation can be observed numerically, and the
transition line between these two states has the qualitative form in
the $D/r$ -- $k$ phase plane indicated in Fig.~1.  Our exact analysis
also provides a basis for a scaling description of this transition.
For the total number of particles, we make the following scaling
ansatz,
$$
N(t)\sim t^\beta\Phi(\epsilon t^\alpha),\eqnoi
$$
where $\epsilon(k)$ is the deviation of $D/r$ from its critical value.
This parameter will be considered positive in the region where
propagation occurs and negative otherwise.  From the fact that $N(t)$
grows linearly with time for $\epsilon>0$ and decays as $t^{-1/2}$
otherwise (for the more generic case of parity non-conserving dynamics
or an even number of initial particles and parity conserving dynamics),
one obtains
$$
\eqalign{
\Phi(x)&\sim x^{(1-\beta)/\alpha}\qquad\qquad\ \,\,(x>0),\cr
&\sim |x|^{-(1+2\beta)/(2\alpha)}\qquad(x<0).\cr
}\eqnoi
$$
Since the exact solution shows that $N(t)$ remains bounded as
$t\to\infty$ for $\epsilon=0$, $\beta$ must equal 0.  (As above, we
consider the case where parity is not conserved, or that the initial
number of particles is even for parity conserving dynamics).  Further,
for small values of $D/r$, this ratio appears in the exact solution in
combinations of the form $D\sqrt t /r$, implying that $\alpha=1/2$.
It therefore follows that $N(t)\propto\epsilon^2t$ for small positive
$\epsilon$.  This cannot be checked against our exact solution, since
the $\epsilon>0$ regime cannot be reached when $k=1$.  On the other
hand, for negative $\epsilon$, the behavior $N(t)\propto(|\epsilon|
\sqrt t)^{-1}$ is predicted, in agreement with \back3.

\chapter{Factorization of the Multiparticle Rate Equations}%

As mentioned in Section 2, the single-particle reaction-diffusion
equation cannot account for the transition between extinction and
propagation unless there is a sign change in the coefficient of the
linear term in the concentration.  Although such a sign change can be
justified heuristically, it is worthwhile to present a systematic
continuum approach which leads to this mechanism.  Our approximate
treatment of the multiparticle rate equations accomplishes this task.
This approach also has the advantage that it can be applied
straightforwardly to different microscopic branching and reaction
mechanisms.  In contrast, while the exact solution of the last section
provides a complete description of the transition in a special case,
this method is neither generalizable nor physically intuitive.  Given
the nature of the approximation in the multiparticle rate equations, we
anticipate that our results will not depend quantitatively on the number
of offspring in a branching event (except for parity-specific features),
but rather, will be generic to the transition between extinction and
propagation.

For simplicity, we study the 1-offspring
BAW in one dimension which can be represented as
$$
A\to2A\qquad A+A\to 0.\eqnoi
$$
It should be mentioned that the 1-offspring BAW with a unit reaction
probability has been investigated by rigorous mathematical
techniques.\ref{1}~ These approaches have established the existence of a
transition between propagation and extinction and provided weak bounds
for the critical value of $D/r$.  Our approximate method locates the
transition for all values of the reaction probability.  We first determine
the first two in the hierarchy of rate equations for the multiparticle
densities in this process.  These equations will be closed by
factorizing three-particle densities as products of two-particle
densities.  To write the hierarchy of rate equations, define
$\rho_{\{k\}}$ as the probability that the set $\{k\}$ is occupied.
Thus $\rho_{0}$ is the probability that site 0 is occupied, $\rho_{0,i}$
is the probability that sites 0 and $i$ are simultaneously occupied,
$\rho_{0,i,j}$ is the probability that 0, $i$, and $j$ are
simultaneously occupied, \etc.  The rate equation for $\rho_0$ is found
by enumerating all configurations in which an elemental event changes
the occupancy of site 0.  For example, if site 0 is empty and site 1 is
occupied (which occurs with probability $\rho_0-\rho_{0,1}$), then by
either branching or hopping to the left, site 0 can become occupied.
There is a similar contribution if site $-1$ is occupied.  Similarly,
there are three elemental events which lead to a decrease in the
occupancy of site 0, as illustrated in Fig.~4(a).  Summing these
contributions, leads to the rate equation
$$
\dot\corrone{0}=r\rho_0-[r+(2D+r)k]\corrtwo{0}{1}\eqnai
$$

Similarly, the rate equation for $\rho_{0,1}$ is obtained by enumerating
all 3-site configurations for which an elemental event changes the
occupancy of sites 0 or 1.  These configurations and the associated
rates of the processes which change $\rho_{0,1}$ are shown in Fig.~4(b)
and lead to
$$
\dot\rho_{0,1}=
r\rho_0-(D+r)(1+k)\rho_{0,1}+(D+r)\rho_{0,2}-[r+(2D+r)k]\rho_{0,1,2}
\eqno(\the\secnum.\the\eqnum{b})
$$
The equations for $\rho_{0,i}$ for $i\ge 2$ are obtained similarly (Fig.~4(c)),
$$
\dot\rho_{0,i}=
(D+r)\rho_{0,i-1}-2D\rho_{0,i}+(D+r)\rho_{0,i+1}-
[r+(2D+r)k](\rho_{0,i,i+1}+\rho_{0,i-1,i}).
\eqno(\the\secnum.\the\eqnum{c})
$$
To be soluble, these exact equations must now be closed by a suitable
truncation.  A standard approach is to truncate the equations at the
2-particle level by the ansatz
$$
\corrthree{0}{1}{i+1}=\corrtwo{0}{1}\corrtwo{1}{i+1}/\corrone{0}=
\corrtwo{0}{1}\corrtwo{0}{i}/\corrone{0}.\eqnoi
$$
While heuristic, this approximation yields reasonable results and turns
out to be simpler than the standard Kirkwood truncation.

To establish the existence of propagation, we investigate the
existence of a non-zero time-independent solution to the truncated rate
equations.  The transition between propagation and extinction is then
identified by the locus where this steady-state density
vanishes as  $k$ and $D/r$ are varied.  To streamline the
notation, let $x\equiv D/r$ and $\rho(i)\equiv\corrtwo{0}{i}/\rho_0$.
The steady-state rate equations reduce to
$$
\eqalign{
1&-[1+(1+2x)k]\rho(1)=0,\cr
1&-(1+x)(1+k)\rho(1)+(1+x)\rho(2)-[1+(1+2x)k]\rho(1)^2=0,\cr
(1&+x)\rho(i-1)-2x\rho(i)+(1+x)\rho(i+1)-[1+(1+2x)k]\rho(1)
(\rho(i)+\rho(i-1))=0.\cr}
\eqnoi
$$
{}From the first of these equations,
$$
\rho(1)={1\over{1+(1+2x)k}}.\eqnoi
$$
Using this in the second equation then gives
$$
\rho(2)={1+x-xk\over{(1+k+2xk)(1+x)}}.\eqnoi
$$
Substituting the static value for $\rho(1)$ into the last of the
rate equations, a constant coefficient recursion equation for
$\rho(i)$ results.  Using the boundary conditions supplied by the
equations for $\rho(1)$ and $\rho(2)$, the solution is,
$$
\rho(i)=\rho+{xk\over{1+k+2xk}}\left({x\over{1+x}}\right)^{i-1},\eqnoi
$$
with
$$
\rho={1-xk\over{1+k+2xk}}.\eqnoi
$$

Here $\rho=\lim_{i\to\infty}\rho(i)$ is just the equilibrium
single-particle density.  Thus a positive solution for $\rho$ exists
only when $x<1/k$, corresponding to the propagating phase of the
1-offspring BAW.  When $x=x_c=1/k$, the equilibrium density vanishes;
this  defines the phase boundary between propagation and
extinction.  Note further that the decay of the equilibrium
concentration is linear in $\epsilon$ as $\epsilon\equiv x-x_c\to 0^+$.  If
the scaling predictions of the preceding section are correct, the
particle number should grow as $\epsilon^2t$, for $\epsilon\to 0^-$.  On
the other hand, the total particle number is the particle density
multiplied by the width of the propagating wave.  This latter quantity
should grow as $\epsilon t$, that is, the front velocity goes linearly
to zero as the critical value of $x$ is approached.

\chapter{Numerical Simulations}%

To confirm the above analytical results and visualize the evolution of
the system, we have performed numerical simulations of the 1-
and 2-offspring BAW in one dimension according to the rules outlined
in Section 2.  For concreteness, we have fixed the hopping rate to be
unity so that the variables in the simulation are the branching rate
$r$ and the reaction probability $k$.

First consider the 1-offspring BAW for which it is known\ref{1,2} that a
transition between propagation and extinction occurs for a non-zero
value of $D/r$.  Starting with a single particle at the origin and with
the parameterizations of our lattice model, the transition occurs at
$x\equiv D/r\cong 1/.89$.  For slightly larger values of $x$, the number
of particles initially increases, but ultimately decays to zero.  For
example, for $x=1/.88$, the average number of particles gradually
increases to 4.8 for $t\simeq 400$, but then decreases to 0 for longer
times.  By $t=1.5^{23}\simeq 11223$, only about $.1\%$ of the initial
configurations are still active.  For $400\ltwid t\ltwid 10,000$, the
spatial distribution of the ensemble of surviving particles appears
visually to be well approximated by a Gaussian (Fig.~5(a)).  In this
time range, the reduced moments of the spatial distribution are
$m_4\equiv \av{x^4}/\av{x^2}^2\cong 2.8$ -- $3.02$ and $m_6\equiv
\av{x^6}/\av{x^2}^3\cong 13.0$ -- $13.9$, while for a Gaussian, the
corresponding values are $m_4=3$ and $m_6=15$.  This behavior suggests
that the effect of branching is irrelevant in the scaling sense, and that
the spatial evolution of the 1-offspring BAW coincides with that of a
single purely random walk.  On the other hand for $x=1/.95$, the simulations
clearly show that a Fisher-like wavefront forms which then propagates at
a finite velocity (Fig.~5(b)).

For the 2-offspring BAW, we have approximately mapped out the phase
boundary between extinction and propagation (Fig.~1(b)).  Simulations
clearly indicate that the phase boundary intersects the line $k=1$ at
$D/r=0$.  Thus for $k=1$ there is extinction for all $D/r$ except when
$D/r=0$.  At this special point, our exact solution showed that the
number of particles grows as $\sqrt{t}$ for a single particle initial
state.  Another amusing feature of the 2-offspring BAW that we have
verified numerically is the essential dependence on parity.  For
example, if the initial state consists of two widely separated
particles, then propagation associated with two independent particles is
observed for early times, followed ultimately by a decay in which the
number of particle goes to zero.

Finally, in the course of simulations we discovered an interesting
difference between serial and parallel updating.  When all lattice
sites are updated simultaneously, the 2-offspring BAW exhibits
propagation, that is, a Fisher wave, even for finite values of $D/r$.
Thus, the BAW process is an example of a system where the method of
updating, whether parallel or serial, leads to non-trivial differences
in the resulting dynamics.  We chose to focus on serial updating,
since it is more physically motivated.

\chapter{Conclusions}%

In summary, we have analyzed the dynamical behavior of branching
annihilating random walks in one dimension.  Typically this system
exhibits (i) propagation, where a localized population of
particles spreads ballistically for large branching and/or small
reaction rates, and (ii) extinction, where a localized population
eventually disappears for the complementary range of parameters.  For
parity non-conserving rules, there is a continuous transition between
these two behaviors which occurs at a finite value of the branching rate
when the reaction probability is unity.  This transition appears to have
a universal character and can be accounted for by analysis of a
truncated hierarchy of rate equations for multiparticle distribution
functions, as well as by scaling.

When the branching and reaction mechanisms conserve parity, rather
different behavior arises.  In particular, for the 2-offspring BAW with
a unit reaction probability, an exact analysis of the master equation
for the parity of the particle number in a fixed length interval shows
that the transition between propagation and extinction occurs for
infinite branching rate.  The long-time behavior also depends
fundamentally on the parity of the initial number of particles.  When
this number is odd, extinction corresponds to a long time state where
the particle number is bounded, while the particle number vanishes as
$t^{-1/2}$ for an even number of initial particles, \ie, extinction is
complete.  In the limit of infinite branching rate, a single initial
particle gives rise to a Gaussian-like density distribution where the
particle number grows as $t^{1/2}$.  Numerical studies confirm many of
these analytical predictions.
\vfill\eject
\bigskip\centerline{\bf Acknowledgements}\medskip

We gratefully acknowledge DGAPA project IN100491, and project \#903922
from CONACYT (FL) and ARO grant DAAL04-93-G-0021 and NSF grant
INT-8815438 (SR), for partial financial support of this research.

\bigskip\bigskip
{\bf References}

\refi M. Bramson and L. Gray, \zw 68 447 1985 .

\refi A. Sudbury, \annpr 18 581 1990 .

\refi H. Takayasu and A. Yu. Tretyakov, \prl 68 3060 1992 .

\refi I. Jensen, \prl 70 1465 1993 .

\refi For general reviews on interacting particle systems, see \eg, T.
M. Liggett, {\sl Interacting Particle Systems}, (Springer-Verlag, New
York, 1985); R. Durrett, {\sl Lecture Notes on Particle Systems and
Percolation}, (Wadsworth, Pacific Grove, CA, 1988).

\refi T. E. Harris, \annpr 2 969 1974 .

\refi F. Schl\"ogl, \zpb 253 147 1972 ; P. Grassberger and A. de la
Torre, \annp 122 373 1979 .

\refi R. C. Brower, M. A. Furman, and M. Moshe, \plb 76 231 1978 ; J. L.
Cardy and R. L. Sugar, \jpa 13 L423 1980 .

\refi R. A. Fisher, \ae 7 353 1937 ;
A. Kolmogoroff, I. Petrovsky and N. Piscounoff, \mubm 1 1 1937 .

\refi J. D. Murray, {\sl Mathematical Biology}, (Springer-Verlag,
Berlin, 1989).

\refi P. Krapivsky, \pra 45 1067 1992 ; \jpa 25 5831 1992 .

\refi H. Takayasu and H. Inui, \jpa 25 L585 1992 .

\vfill\eject

{\bf Figure Captions}
\medskip

\rfigi Phase diagram of the BAW process as a function of the reaction
probability $k$ and the ratio of the diffusion to branching rate $D/r$.
In (a) the phase diagram appropriate for BAW processes with non-parity
conserving branching is indicated.  This behavior should be taken as
representative of the continuum limit of the BAW process.  In (b), the
phase diagram for the 2-offspring BAW is schematically indicated.  In
this case, the nature of the ``extinct'' phase depends on whether the
initial number of particles is even or odd.

\rfigi Illustration of the space-time evolution of the 2-offspring BAW
in one dimension in the pure branching limit, $D/r=0$, and $k=1$.  In
this case, an isolated nearest-neighbor pair of particles remains bound
and executes a simple random walk.  The particle which branches at the
next step is indicated by the open circle.

\rfigi Enumeration of the processes which can change the parity of the
number of particles contained in the interval $[j,k]$ under the action
of the 2-offspring BAW.  Shown are the various microscopic events and
their corresponding statistical weights.  The cases of (a) $k>j$ and (b)
$k=j$ are somewhat different and therefore need to be treated separately.

\rfigi Enumeration of the processes which can change the occupancy of
a selected set of sites in the 1-offspring BAW.  In (a) the set consists
of a single site, $\rho_{0}$, while in (b) the set consists of 2 adjacent
sites,
$\rho_{0,1}$.  In
(c), the set consists of 2 non-adjacent sites, $\rho_{0,i}$.  The signs to
the left
denote whether a process increases ($+$) or decreases ($-$) $\rho$.
The rate of the processes (and their kind) is indicated to the right. Solid
and empty circles denote occupied and empty sites, respectively.

\rfigi Spatial distribution for the 1-offspring BAW in one dimension at
$t=1.5^{19}\cong 2217$.  The cases shown are: (a) 25000 configurations at
$D/r=1/.88$, where the survival probability has decreased to
approximately 0.11, and (b) 200 configurations at $D/r=1/.95$, where the
survival probability has saturated at approximately 0.81.

\bye